\documentclass[12pt,letterpaper]{article}
\usepackage[a4paper, total={7in, 10in}]{geometry}

\usepackage{graphicx}
\usepackage{helvet}
\usepackage{authblk}
\usepackage{hyperref}
\usepackage{amsmath} 
\usepackage{amssymb} 
\usepackage{orcidlink} 
\usepackage[super,comma,sort&compress]  
   {natbib}

\makeatletter
\renewcommand{\maketitle}{\bgroup\setlength{\parindent}{0pt}
\begin{flushleft}
  \textbf{\@title}
  
  \@author
\end{flushleft}\egroup}
\makeatother

\newcommand{\ket}[1]{| #1 \rangle}

\begin{document}

\title{Photonic Hybrid Quantum Computing}
\date{}

\author[1,2,\orcidlink{0000-0002-3011-1129}]{Jaehak Lee}
\author[3,\orcidlink{0000-0002-3112-9101}]{Srikrishna Omkar}
\author[3,\orcidlink{0000-0002-1766-6402}]{Yong Siah Teo}
\author[4,\orcidlink{0000-0002-1207-2752}]{Seok-Hyung Lee}
\author[5,\orcidlink{0000-0001-5520-0905}]{Hyukjoon Kwon}
\author[6,5,*,\orcidlink{0000-0001-5543-5050}]{M. S. Kim}
\author[3,**,\orcidlink{0000-0003-0348-3397}]{Hyunseok Jeong}

\affil[1]{Center for Quantum Technology, Korea Institute of Science and Technology (KIST), Seoul 02792, Korea}
\affil[2]{Division of Quantum Information, KIST School, Korea University of Science and Technology (UST), Seoul 02792, Korea}
\affil[3]{NextQuantum Innovation Research Center, Department of Physics and Astronomy, Seoul National University, Seoul 08826, Korea}
\affil[4]{Centre for Engineered Quantum Systems, School of Physics, The University of Sydney, Sydney, NSW 2006, Australia}
\affil[5]{Korea Institute for Advanced Study, Seoul 02455, Korea}
\affil[6]{Blackett Laboratory, Imperial College London, London SW7 2AZ, United Kingdom}

\affil[*]{m.kim@imperial.ac.uk}
\affil[**]{jeongh@snu.ac.kr}

\maketitle

\section*{SUMMARY}

Photons are a ubiquitous carrier of quantum information: they are fast, suffer minimal decoherence, and do not require huge cryogenic facilities. Nevertheless, their intrinsically weak photon–photon interactions remain a key obstacle to scalable quantum computing. 
This review surveys hybrid photonic quantum computing, which exploits multiple photonic degrees of freedom to combine the complementary strengths of discrete and bosonic encodings, thereby significantly mitigating the challenge of weak photon–photon interactions.
We first outline the basic principles of discrete-variable, native continuous-variable, and bosonic-encoding paradigms. We then summarise recent theoretical advances and state-of-the-art experimental demonstrations with particular emphasis on the hybrid approach. 
Its unique advantages, such as efficient generation of resource states and nearly ballistic (active-feedforward-free) operations, are highlighted alongside remaining technical challenges.
To facilitate a clear comparison, we explicitly present the error thresholds and resource overheads required for fault-tolerant quantum computing.
Our work offers a focused overview that clarifies how the hybrid approach enables scalable and compatible architectures for quantum computing.

\section*{KEYWORDS}

Optical quantum computing, Hybrid entanglement, Fault-tolerance

\section*{Introduction}

Quantum computing has made remarkable progress across various physical platforms. 
Recent advances in superconducting circuits, ion traps, and neutral atoms have significantly improved the accuracy of gate operations. However, it is widely anticipated that these platforms will face limitations when it comes to scaling the number of qubits to the level required for fault-tolerant quantum computing (FTQC). There is still a huge effort required to precisely prepare, control and measure a large quantum system at a fast gate speed. In fact, the number of controllable qubits required for FTQC with error correction is generally expected to be orders of magnitude greater than what is reachable in the foreseeable future. One possible route to overcome some of these limitations is through hybrid computing with photonic qubits, which offer fast operation and low decoherence.

Photonic quantum computing based on discrete variables (DV) typically relies on single-photon qubits and linear optical elements \cite{Knill2001a,Kok2007a}. 
A major challenge has been the inherently nondeterministic nature of this approach
\cite{Lutkenhaus1999a,Calsamiglia2001a}. Various efforts have been made to address this issue using prearranged offline entangled states \cite{Gottesman1999a,Raussendorf2001a}. Recent theoretical progress has significantly improved the loss thresholds for single-photon-based quantum computing
\cite{Omkar2022a,Lee2023a,Bartolucci2023a,Song2024a,Bartolucci2025}. However, practical implementation remains still highly challenging due to the large number of physical qubits required for error correction or the continual need for active feedforward operations during gate operations.

Continuous-variable (CV) quantum computing exploits the continuous degrees of freedom of the infinite-dimensional Hilbert space \cite{Lloyd1999a, Braunstein2005a, Menicucci2006a, Gu2009a, Marek2011a}. 
It utilizes CVs as information carriers, called qumodes, instead of discrete qubits or qudits. 
However, while finite universal sets of CV gates are known, 
a complete fault-tolerant model for quantum computing based solely on universal CV gates remains to be established \cite{Blair2025a}.

A bosonic qubit is a logical two-level quantum system encoded within the infinite-dimensional Hilbert space of a single bosonic mode \cite{Albert2018a}. 
It is constructed by selecting two (approximately) orthogonal states that serve as the logical basis.
An example of such a scenario is the Gottesman-Kitaev-Preskill (GKP) states \cite{Gottesman2001a}.
Quantum computing  based on the GKP states \cite{Gottesman2001a} has attracted considerable attention because of its potential for relatively efficient error correction. The main difficulty, however, lies in the fact that generating high-quality photonic GKP states efficiently is extremely challenging, and no suitable method has yet been firmly established.
The binomial code utilizes superpositions of Fock states with carefully designed binomial coefficients to protect against specific errors \cite{Michael2016a}. 
While it offers a robust structure for error correction against photon loss and dephasing using orthogonal superpositions of finite Fock states, 
the high complexity of resource state preparation and sensitivity to realistic, compounded noise limit its practical scalability \cite{Albert2018a}.

Another bosonic qubit is a superposition of two coherent states, commonly referred to as a cat state \cite{Cochrane1999a, Jeong2002a, Ralph2003a}, based on which all-optical quantum computing schemes were suggested \cite{Jeong2002a, Ralph2003a}.
Cat-state qubits offer distinct advantages in the context of all-optical implementations. They enable a nearly deterministic Bell-state measurement (BSM) using a simple optical setup consisting of a beam splitter and two photodetectors \cite{Jeong2001a,Jeong2002b} — a feat not achievable with single-photon qubits. This simplicity facilitates highly efficient gate operations, significantly enhancing resource efficiency \cite{Ralph2010a}.
Furthrmore, several methods for generating cat states have been suggested and experimentally demonstrated \cite{Ourjoumtsev2006a, Ourjoumtsev2007a}. Nonetheless, cat-state qubits are composed of coherent states that are inherently non-orthogonal, which makes certain single-qubit operations difficult unless sufficiently large cat states are available. Additionally, due to their intrinsic property of having undefined photon number, cat states tend to be relatively vulnerable to photon loss.

There have been efforts to develop more powerful schemes for quantum computing by hybridization of different types of approaches \cite{Andersen2015a}.
Photonic hybrid quantum computing, where a logical qubit is an entangled combination of a single-photon part and a coherent-state part \cite{Lee2013a, Omkar2020a, Omkar2021a, Lee2024a}, combines the strengths of the above methods in a highly synergistic way.
Unlike GKP states, hybrid entangled states can be generated more efficiently \cite{Jeong2014a, Morin2014a}, and experimental demonstrations have already been reported \cite{Jeong2014a, Morin2014a, Ulanov2017a, Sychev2018a, Guccione2020a, Darras2023a}. Like cat-state-based approaches, hybrid methods allow for deterministic BSMs \cite{Lee2013a}. Furthermore, the DV component of the hybrid qubit provides an orthogonal basis regardless of the amplitude of the coherent-state part, which is distinct from a typical cat-state qubit. As a result, hybrid architectures can simultaneously achieve a reasonably high level of fault-tolerance and resource efficiency, while operating in an almost entirely ballistic manner that avoids most of active feedforward processes. These features make hybrid photonic quantum computing a highly promising and competitive platform in the long-term pursuit of scalable quantum computing technologies. 
In this topical review, we report on the recent progress and future prospects of the hybrid methods for photonic quantum computing.

\section*{Discrete-variable approach}

Knill, Laflamme, and Milburn (KLM) proposed a protocol for scalable linear optical quantum computing (LOQC), using only linear optical elements, single-photon sources, and photon-counting detectors \cite{Knill2001a}. 
In this framework, qubits are defined via dual-rail encoding \cite{Kok2007a}, where logical states $\ket{0}$ and $\ket{1}$ correspond to a single photon occupying one of two orthogonal modes (e.g., two distinct polarizations or spatial paths).
The KLM protocol leverages measurement-induced nonlinearities to implement non-deterministic 
entangling gates via teleportation and post-selection \cite{Gottesman1999a}, overcoming a fundamental challenge in photonic quantum computing, that is, the absence of strong interactions between photons.

\begin{figure}
    \centering
    \includegraphics[width=0.9\columnwidth]{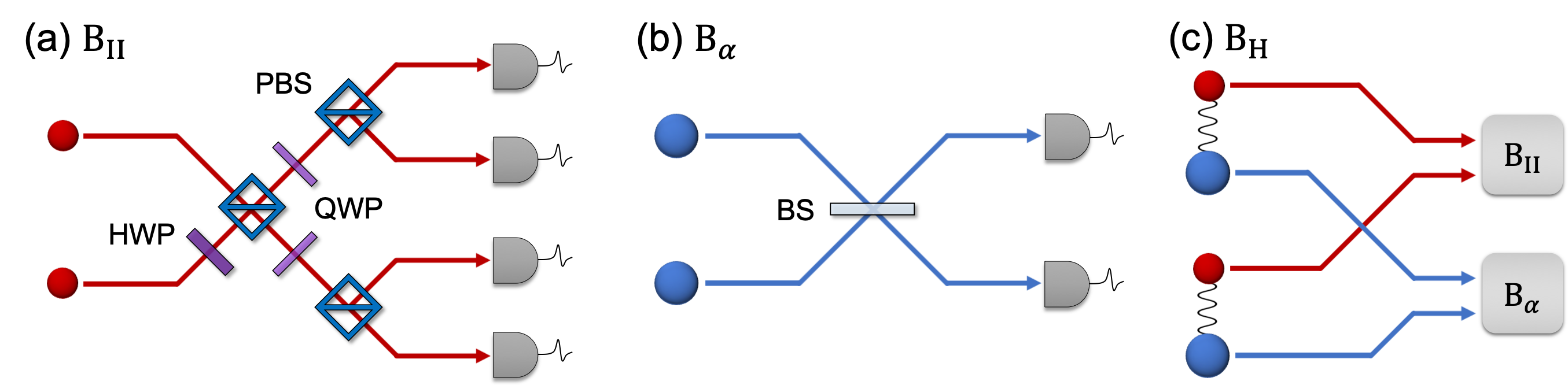}
    \caption{BSM schemes for three types of encoding. (a) DV BSM, $ \textrm{B}_\textrm{II} $, consisting of polarizing beam splitters (PBSs), a half waveplate (HWP), two quarter waveplates (QWPs), and four on-off detectors. (b) BSM in coherent-state basis, $ \textrm{B}_\alpha $, implemented by a beam splitter (BS) and two PNR detectors. (c) HBSM, $ \textrm{B}_\textrm{H} $, realized by separately performing two BSMs $ \textrm{B}_\textrm{I} $ and $ \textrm{B}_\alpha $. The red (blue) circles represent single-photon (cat-state) qubits. While the success probability of  $ \textrm{B}_\textrm{II} $  is 1/2, those of $ \textrm{B}_\alpha $ and  $ \textrm{B}_\textrm{H} $ can be made arbitrarily high by increasing coherent amplitude $\alpha$. }
    \label{fig:Bell}
\end{figure}

As used in the KLM protocol, teleportation 
is a key technique for addressing the non-determinicity of entangling gates, as it allows shifting a large portion of probabilistic operations into offline preparation of resource states \cite{Gottesman1999a}.
However, consuming these states to implement entangling gates involves BSMs, which remain inherently probabilistic.
Under standard LOQC settings, the success probability of a BSM cannot exceed 50\% without additional resources \cite{Lutkenhaus1999a, Calsamiglia2001a}.
A BSM scheme, also called as type-II fusion operation \cite{Browne2005a}, is illustrated in Fig.~\ref{fig:Bell}(a), which can 
distinguish only two among the four Bell states.
It is thus crucial to enhance the success probability of a BSM for scalability of LOQC. 
Methods have been proposed for this purpose employing squeezing operations \cite{Zaidi2013a}, a pair of ancillary entangled photons \cite{Grice2011a}, or unentangled ancillary single photons \cite{Ewert2014a}, reaching a success probability of 75\% \cite{Grice2011a,Ewert2014a}.
It was shown that a logical BSM can be performed  with arbitrarily high success probability of $1 - 2^{-N}$ using $N$ multi-photon encoding and $N$ physical BSMs  \cite{Lee2015a}.

LOQC schemes using measurement-based nonlinearities and teleportation have further advanced through the integration of measurement-based quantum computing (MBQC) \cite{Raussendorf2001a}, a paradigm that performs computation solely by single-qubit destructive measurements on a large entangled \emph{cluster state}. 
Given a graph $G$, a cluster state $\ket{G}$ is defined as a state that can be prepared by initializing each qubit to $\ket{+}$ and applying a CZ gate on each pair of qubits connected by an edge.
MBQC can be viewed as an extension of quantum teleportation:
Input qubits that encode the initial state are first entangled with a cluster state, followed by single-qubit measurements on all qubits except a designated subset of ``output qubits'', which effectively teleports the original input state onto the output qubits with a feedforward operation.
MBQC is equivalent to conventional circuit-based quantum computing and thus capable of universal quantum computing \cite{Raussendorf2001a}.   

MBQC is particularly advantageous for LOQC because non-deterministic operations are, in principle, only required during the offline preparation of cluster states, while the primary computation itself relies on straightforward single-qubit measurements. 
Photonic cluster states can be generated using type-I or -II ``fusion'' operations, $ \textrm{B}_\textrm{I} $ or $ \textrm{B}_\textrm{II} $ \cite{Browne2005a}, where the latter corresponds essentially to a destructive BSM.
Starting with small ``building-block'' cluster states (e.g., three-photon Greenberger-Horne-Zeilinger state (GHZ) states), one can iteratively merge them via fusions to construct the desired cluster state.

Due to the limited success probabilities of fusion operations and various noise sources such as photon loss, robust QEC codes are essential to ensure robustness against errors.
QEC codes enable arbitrarily large computation with small error rates, when the physical error rates remain below a certain limit, called a \emph{noise threshold}.
In this context, a scheme using concatenated QEC codes and parallel fusions has demonstrated clear noise thresholds \cite{Dawson2006a}.
Additionally, topological QEC codes (particularly surface codes) integrated naturally into MBQC using a family of three-dimensional cluster states, called the Raussendorf-Harrington–Goyal (RHG) lattice \cite{Raussendorf2006a}.
This has been adapted to photonic qubits by incorporating fusions, demonstrating thresholds against both photon loss and computational errors \cite{Herrera-Marti2010a}. 
Although the photon loss thresholds obtained by initial investigations were low ($\sim$ 0.05\% \cite{Herrera-Marti2010a}), recent advances have significantly improved them to practically relevant levels up to about 10\% in a more resource-efficient way than previous schemes \cite{Omkar2022a,Lee2023a}.
These improvements utilize near-deterministic logical-level fusions with multi-photon encoding, enabling end-to-end schemes from 3-photon GHZ states to fault-tolerant MBQC  \cite{Omkar2022a,Lee2023a}.
Another possible direction is ``ballistic'' photonic MBQC \cite{GimenoSegovia2015a,Herr2018a}, which avoids post-selection by employing graph purification and thereby reduces the technical complexity associated with active switching and feedforward operations. However, its impact on error thresholds and resource requirements may vary depending on the underlying architecture.

Fusion-based quantum computing (FBQC) \cite{Bartolucci2023a}  is a variant of the LOQC framework that inverts the relationship between fusions and measurements compared to MBQC.
While MBQC uses fusions to prepare large entangled resource states for subsequent single-qubit measurements, FBQC employs fusions themselves (between constant-size resource states) as the primary computational mechanism.
This approach may enable more streamlined and flexible LOQC strategies by eliminating the intermediate cluster state construction step.
Similarly to MBQC, key challenges in FBQC include photon loss and the inherently nondeterministic nature of fusion operations, implying that many techniques developed for MBQC can be directly adapted to FBQC. 
For instance, encoding resource states using quantum parity codes \cite{Ralph2010a,Hayes2010a} can lead to a photon loss threshold exceeding 10\% \cite{Song2024a,Bartolucci2025}, while increasingly heavy resource requirements are still an obstacle.
Additionally, there are ongoing efforts to generalize FBQC by using GHZ-state measurements involving more than two qubits \cite{Pankovich2024a}, potentially leading to higher loss tolerance compared to standard FBQC at the cost of increased technical complexity.

\section*{Continuous-variable approach}

CV quantum information processing exploits continuous degrees of freedom inherent in the infinite-dimensional Hilbert space of a quantum harmonic oscillator. The CV approach represents and manipulates the unit of information (``qunat") directly using physical quantities with CVs such as $x$ and $p$ quadratures. 
A significant milestone in CV quantum computation is the realization of large-scale squeezed-state cluster states, which serve as a universal resource for CV MBQC \cite{Menicucci2006a}. Recent experiments have demonstrated the generation of two-dimensional CV cluster states by time-domain multiplexing of squeezed states using optical delay loops followed by beam-splitter interactions \cite{Yokoyama2013a, Asavanant2019a, Larsen2019a, Larsen2021a}. At each time bin, a two-mode squeezed state is produced by interfering two squeezed states on a beam splitter. Two-mode squeezed states are then multiplexed using optical delay lines, effectively linking CV modes at different time bins, which in turn connects the CV modes into a two-dimensional cluster state.

A key advantage of this approach lies in the deterministic nature of Gaussian operations, described by Hamiltonians which are second-order polynomials in quadrature operators, allowing for the scalable and efficient generation of  cluster states. Nevertheless, 
since a circuit with Gaussian operations alone can be efficiently simulated by a classical computer \cite{Bartlett2002a}, non-Gaussian resources requiring third- or higher-order nonlinearity must be incorporated. It has been shown that a single non-Gaussian operation, for example, a cubic phase gate 
\cite{Lloyd1999a} suffices to achieve universal CV quantum computation \cite{Lloyd1999a}. However, implementing such a gate with arbitrary nonlinearity remains technically challenging. As an alternative,
by injecting non-Gaussian resource states, prepared offline, into a 
CV cluster state, non-Gaussian gates can be implemented via gate teleportation
\cite{Asavanant2023}.
Another noteworthy method is to introduce a measurement-induced nonlinearity, where non-Gaussian measurements, such as photon-number-resolving (PNR) detection, provide the required non-Gaussianity and enable universal computation \cite{Menicucci2006a}.
Despite ongoing progress \cite{Blair2025a}, a fully fault‑tolerant quantum computing scheme based solely on universal CV gates has not yet been established.

\section*{Quantum computing using bosonic qubits} 

\subsection*{Quantum computing using cat states}

    \begin{figure}
        \centering
        \includegraphics[width=0.7\columnwidth]{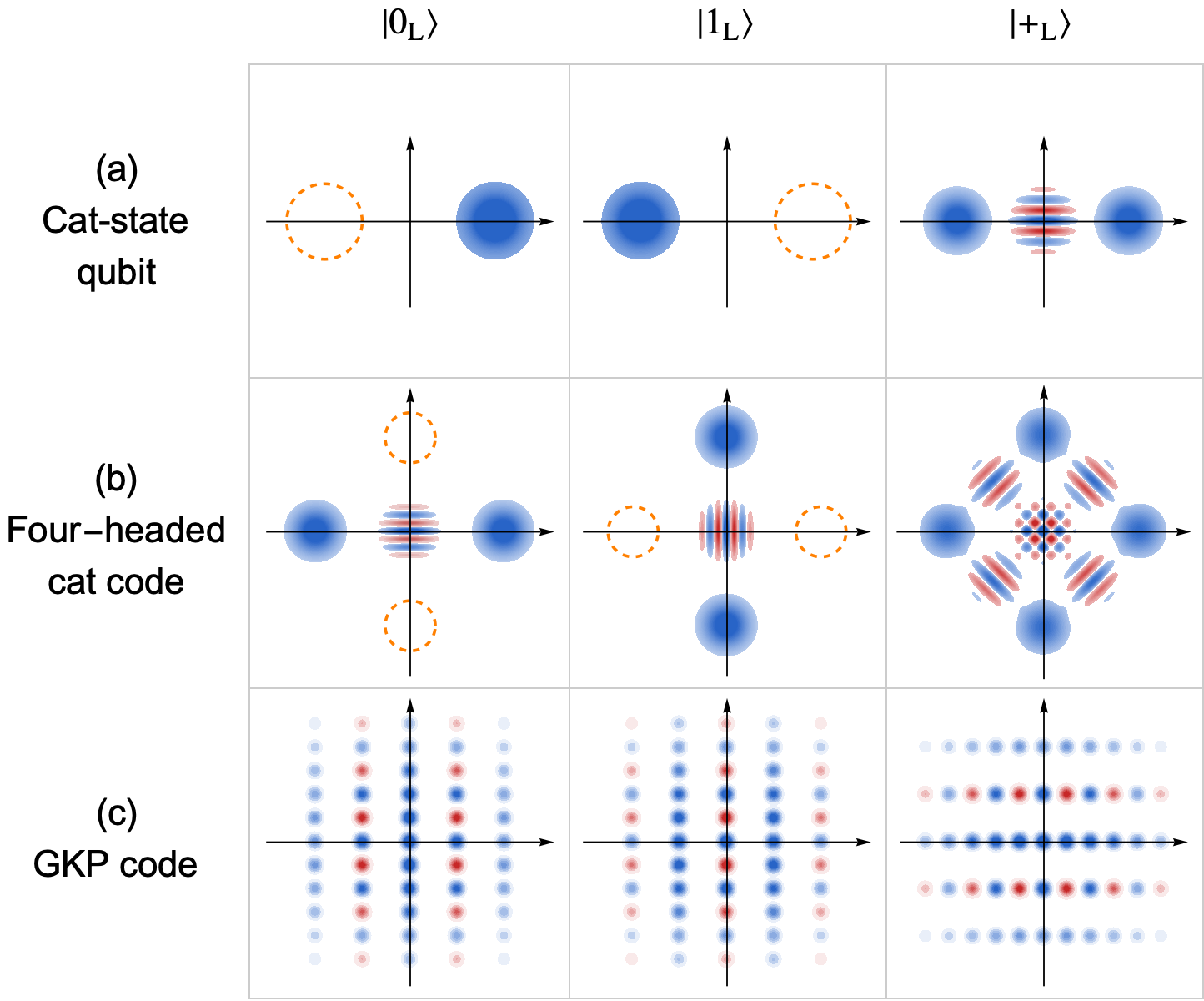}
        \caption{Wigner function of logical states $ \ket{0_\textrm{L}} $,  $ \ket{1_\textrm{L}} $, and $ \ket{+_\textrm{L}} $ for bosonic encoding schemes: (a) cat-state qubit, (b) four-headed cat code, (c) GKP code. Blue (red) regions represent positive (negative) Wigner function and orange circles show unoccupied logical space.}
        \label{fig:BosonicQubit}
    \end{figure}

The photonic schemes for coherent-state quantum computing (CSQC) typically utilize two coherent states, $ \ket{\pm\alpha} $ with amplitudes $\pm\alpha$, as the logical basis \cite{Jeong2002a, Ralph2003a}. 
Their equal superpositions, $ \ket{\pm_\alpha}\propto \ket{\alpha}\pm \ket{-\alpha} $, are called even and odd cat states, due to their definite photon-number parities, respectively.
One of the most significant advantages of CSQC, in contrast to LOQC, is the ability to perform near-deterministic BSMs \cite{Jeong2001a,Jeong2002b}. The scheme for BSM in the coherent-state basis is presented in Fig. \ref{fig:Bell}(b), where a 50:50 beam-splitter interaction is followed by photon-number parity measurements. 
After passing through the beam splitter, the four Bell states,
$\ket{\alpha}_A\ket{\alpha}_B\pm \ket{-\alpha}_A\ket{-\alpha}_B  $ and $  \ket{\alpha}_A\ket{-\alpha}_B\pm \ket{-\alpha}_A\ket{\alpha}_B$, become $\ket{\pm_\alpha}_A\ket{0}_B$ and $\ket{0}_A\ket{\pm_\alpha}_B$, so that they can be unambiguously discriminated.
The BSM fails only when both detectors click no photon, where the failure probability is determined by the overlap between two coherent states, that is $ p_f = e^{-2\alpha^2} $, and rapidly converges to 0 as $ \alpha $ becomes large.
The near-deterministic BSM enables universal quantum computing based on gate teleportation with off-line resource states \cite{Jeong2002a, Ralph2003a}. 
Building on this framework, Ref. \cite{Lund2008a} proposed an error correction scheme based on telecorrection protocol and demonstrated the fault-tolerance of CSQC where the optimized amplitude is $\alpha\approx1.5$.
The CSQC relies on the ability to generate Schr{\" o}dinger cat states, $\ket{\pm_\alpha}$, which serve as fundamental building blocks for the preparation of logical qubit states. One may also construct the logical basis using arbitrary superpositions  of coherent states \cite{Cochrane1999a, Jeong2001a} or using more than two coherent-state components \cite{Leghtas2013a, Mirrahimi2014a}.

Cat states in free-traveling fields have been experimentally generated using schemes based on conditioning measurements \cite{Neergaard-Nielsen2006a, Ourjoumtsev2006a, Ourjoumtsev2007a, Takahashi2008a}. Small cat states ($\alpha\approx1.2$) were demonstrated using single-photon subtraction \cite{Neergaard-Nielsen2006a, Ourjoumtsev2006a} while cat states with clear separation between the coherent states in phase space ($\alpha\approx1.6$) were implemented using conditioning homodyne measurements \cite{Ourjoumtsev2007a} and multiple photon subtractions \cite{Takahashi2008a}.
Further, the breeding protocol enhances the size of cat states by combining smaller cat states to produce larger ones \cite{Lund2004a, Etesse2015a, Sychev2017a}, achieving the amplitude $ \alpha \approx 1.85 $.

\subsection*{Bosonic error correction codes}

Bosonic error correction codes make bosonic qubits resilient to typical errors, mainly due to photon loss. While conventional error correction codes protect a logical information by distributing it across many physical qubits, it is possible to encode a logical qubit within a single harmonic oscillator by exploiting its infinite-dimensional Hilbert space. One of the well-known examples is the binomial code, which encodes qubits in superpositions of Fock states designed to detect errors expressed in terms of annihilation and creation operators \cite{Michael2016a}.

The cat code represents a logical qubit as a superposition of coherent states, with its error resilience enhanced by increasing the number of coherent-state components. The four-headed cat code employs four components of coherent states, encoding a logical qubit in even cat states \cite{Leghtas2013a, Mirrahimi2014a}, specifically $ \ket{0_\textrm{L}} \propto \ket{\alpha} + \ket{-\alpha} $ and  $ \ket{1_\textrm{L}} \propto \ket{i\alpha} + \ket{-i\alpha} $, as shown in Fig.~\ref{fig:BosonicQubit}(b). Within this encoding, a photon loss can be detected by the change in the photon-number parity. In practice, a nonlinear interaction is required to detect photon loss \cite{Bergmann2016a, Grimsmo2020a}, which remains challenging in optical systems. Recent works have proposed schemes for performing BSMs on cat-code qubits, using linear optical elements and PNR detectors, which can simultaneously detect photon loss during measurements \cite{Hastrup2022a, Su2022a}. In addition, a linear optical scheme for generating resource states have been proposed, where a four-component cat state is created by mixing a pair of two-component cat states via a beam splitter \cite{Hastrup2020a}.


Another well-known bosonic error correction code is the GKP code, which encodes a logical qubit into grid-like superpositions in phase space, as shown in Fig.~\ref{fig:BosonicQubit}(c), allowing the correction of small displacement errors \cite{Gottesman2001a}. Due to the high error-correcting performance of the GKP code, architectures based on GKP qubits have been widely investigated towards FTQC \cite{Fukui2018a, Bourassa2021a, Tzitrin2021a}. Once high-quality GKP-state sources are available, a fault-tolerant architecture can be constructed efficiently using only linear optical operations and homodyne measurements \cite{Tzitrin2021a}, and a prototype of this approach has recently been demonstrated \cite{Aghaee-Rad2025a}. However, a major hurdle is that the generation of photonic GKP state is extremely challenging, as it requires strong nonlinearity. Several schemes for generating GKP states have been proposed, particularly based on the breeding protocol \cite{Vasconcelos2010a, Weigand2018a} and the Gaussian boson sampling (GBS) protocol \cite{Tzitrin2020a, Fukui2022a, Takase2023a}.
Both protocols have recently been demonstrated in proof-of-principle experiments \cite{Konno2024a, Larsen2025a}, but the quality and generation rate of the produced states are still far below the requirement for large-scale FTQC.


\section*{Hybrid approach}

One can combine bosonic and DV qubits to take advantages of both degrees of freedom, overcoming the limitations of probabilistic entangling operations for DV qubits and the nonorthogonality of bosonic qubits. In the hybrid approach, we use two physical degrees of freedom, horizontal- and vertical-polarization states of a single photon ($\ket{H}$ and $\ket{V}$) and two coherent states ($ \ket{\pm\alpha}$).  We then define the logical basis of hybrid qubits as $ \ket{0_\textrm{L}} \equiv \ket{+}\ket{\alpha} $ and $ \ket{1_\textrm{L}} \equiv \ket{-}\ket{-\alpha} $, where 
$\ket{\pm} = (\ket{H} \pm \ket{V})/\sqrt{2} $
\cite{Lee2013a}. We can generalize the approach by replacing $\ket{\pm\alpha}$ with other bosonic encodings or DV mode with different degrees of freedom, offering an orthogonal hybrid basis.

\subsection*{Gate operations and teleportation} 

In order to construct a universal set of gate operations, we employ arbitrary $ Z $ rotation ($ Z_\theta $), Pauli $ X $, Hadamard ($ H $), and controlled-$Z$ ($ CZ $) gates. In the hybrid approach, it is possible to implement $ Z_\theta $ and $ X $ gate operations efficiently using liner optics. An arbitrary rotation along  $ Z $-axis is performed by rotating the polarization of DV mode alone. The $ X $ gate requires flipping the polarization of DV mode and applying a $\pi$-phase shift on the bosonic mode. For other operations, namely $ H $ and $ CZ $ gates, we resort to \emph{gate teleportation}. To implement gate teleportation, we require two essential resources, the BSM and the entangled resource state.

The hybrid BSM (HBSM) can be realized by performing two seperate small BSMs, that is, $ \textrm{B}_\textrm{II} $ on DV qubits and $ \textrm{B}_\alpha $ on bosonic qubits, respectively, as shown in Fig. \ref{fig:Bell}(c).
The HBSM fails only if both $ \textrm{B}_\alpha $ and $ \textrm{B}_\textrm{II} $ fail, resulting in a failure probability of $ P_f = \frac{1}{2} e^{-2\alpha^2} $, which is reduced by half compared to the failure probability $ p_f $ of $ \textrm{B}_\alpha $ alone \cite{Lee2013a}.
An advantage of gate teleportation arises from the high success rate of the HBSM. Failures in HBSMs result in logical errors, but these are exponentially suppressed by increasing the encoding amplitude $ \alpha $. Consequently, a FTQC architecture based on this scheme can be more resource-efficient, with a smaller optical footprint due to less switching requirements. However, there is a major bottleneck here; under photon loss a larger encoding amplitude leads to stronger dephasing noise, inducing logical $ Z $ errors. Therefore, it is crucial to find an optimal encoding amplitude that balances this trade-off and maximizes the fault-tolerance threshold.

\subsection*{Hybrid FTQC schemes}

\begin{table*}
    \centering
\begin{tabular}{ c c c c c c }\hline \hline
Scheme & QEC Code  & \begin{tabular}{c} Loss threshold{\textsuperscript{a}} \\ ($\eta_{\rm th}$) \end{tabular} & \begin{tabular}{c} Amplitude \\ ($\alpha_{\rm opt}$) \end{tabular} & \begin{tabular}{c} Resource cost{\textsuperscript{b}} \\ ($N$) \end{tabular} \\
\hline
HQQC \cite{Lee2013a}  & 7-qubit Steane & $4.6\times10^{-4}$  & $1.08$ &$8.2\times10^{9}$\\
HTQC \cite{Omkar2020a} &RHG lattice & $5.07\times10^{-3}$  & $1.50$ & $8.5\times10^5$ \\
PHTQC-2 \cite{Omkar2021a} & RHG lattice & $1\times10^{-2}$ & $0.84$ & $ 1.1\times10^6$ \\
PHTQC-3 \cite{Omkar2021a} & RHG lattice & $ 1.14\times10^{-2}$  & $0.60$ & $2.9\times10^7$ \\
HCQC \cite{Lee2024a} & RHG lattice & $8.9\times 10^{-3}$ & $2.93$  & $2.7\times 10^{4}$ \\  
\hline \hline \end{tabular} 
 \caption{The table lists several hybrid optical QC schemes \cite{Lee2013a,Omkar2020a,Omkar2021a,Lee2024a}, QEC codes associated with them, type of optical resource used, optimal photon-loss thresholds ($\eta_{\rm th}$), coherent-state amplitudes for optimal performance ($\alpha_{\rm opt}$) and incurred resource overheads ($N$).   Apparently, PHTQC-2 and PHTQC-3 offer higher loss thresholds with small amplitudes, while HCQC significantly reduces the resource overhead at the cost of a larger optimal encoding amplitude. 
The PHTQC-$n$ requires hybrid states with small values of $\alpha_{\rm opt}$ which eases resource-state generation. \textsuperscript{a}The referenced publications follow different conventions to quote their loss thresholds.  While HQQC, HTQC, PHTQC-$n$ refer to loss rate of Bell-measurement outcomes, HCQC mentions loss rate of photons. Moreover, HQQC and HCQC refer to the total loss rate, $\eta_{\rm th}$, i.e., an effective rate that compounds loss rates of preparation ($\eta_p)$, storage ($\eta_s)$, gate ($\eta_g)$, and measurement ($\eta_m)$. Therefore, $\eta_{\rm th}=1-(1-\eta_{\rm p})(1-\eta_{\rm s})(1-\eta_{\rm g})(1-\eta_{\rm m})$. On the other hand, HTQC and PHTQC-$n$ refers to thresholds in terms of individual loss process, where $\eta_{\rm p}=\eta_{\rm s}=\eta_{\rm g}=\eta_{\rm m}$. In this table, to make a fair and consistent comparison, we have carefully re-expressed the loss thresholds  $\eta_{\rm th}$ in terms of {\it total photon-loss rate per photonic mode} as well as all associated values. 
 \textsuperscript{b}The resource cost  $N$ is the number of hybrid pairs required to reach a target logical error rate of $p_{\rm L}\sim10^{-6}$ when schemes operate below (about half) the thresholds.
} 
\label{tab:para}
\end{table*}

Photon loss occurs during storage, gate operations, and measurement, introducing logical errors on hybrid qubits. We characterize the total loss rate from resource state preparation to measurement by parameter $ \eta $. Photon loss increases the failure probability $ P_f $ of HBSM and induces logical $ Z $ errors, which can be characterized in terms of $ \eta $ and $ \alpha $. To circumvent these errors, several QEC schemes have been investigated \cite{Lee2013a, Omkar2020a, Omkar2021a, Omkar2022a, Lee2024a}, demonstrating that the hybrid approach can achieve FTQC with reasonable encoding amplitude $ \alpha $.

Table~\ref{tab:para} lists loss thresholds, optimized amplitudes, and  resource costs for several hybrid FTQC schemes. The resource cost $N$ is estimated by counting the number of unit resources (hybrid pairs $ \ket{H}\ket{\alpha} + \ket{V}\ket{-\alpha}$) to achieve a logical error rate below $10^{-6}$. Although a direct comparison between these schemes is not straightforward, as they employ different types of error correction strategies and notations,  we have carefully re-expressed the loss thresholds to make a fair and meaningful comparison from Refs.~\cite{Lee2013a, Omkar2020a, Omkar2021a, Omkar2022a, Lee2024a}. Each loss threshold, $\eta_{\rm th}$, presented in the table and in the following, corresponds to {\it a total photon loss rate per photonic mode}.

The first suggestion for hybrid-qubit-based quantum computation (HQQC) \cite{Lee2013a} introduces a telecorrection-based circuit. In this scheme, the error correction protocol employs several levels of concatenation using the 7-qubit Steane code. The telecorrection circuit is composed of $ H $, $ CZ $, preparation of $ \ket{+}_L $, and $X$-basis measurement. For this purpose, resource states $ \ket{\Phi_H} \propto \ket{0_\textrm{L},0_\textrm{L}} + \ket{0_\textrm{L},1_\textrm{L}} + \ket{1_\textrm{L},0_\textrm{L}} - \ket{1_\textrm{L},1_\textrm{L}} $ and $ \ket{\Phi_{CZ}} \propto \ket{0_\textrm{L},0_\textrm{L},0_\textrm{L},0_\textrm{L}} + \ket{0_\textrm{L},0_\textrm{L},1_\textrm{L},1_\textrm{L}} + \ket{1_\textrm{L},1_\textrm{L},0_\textrm{L},0_\textrm{L}} - \ket{1_\textrm{L},1_\textrm{L},1_\textrm{L},1_\textrm{L}} $ are prepared by merging hybrid pairs and single-photon pairs using $ \textrm{B}_\textrm{I} $ \cite{Browne2005a}  and $ \textrm{B}_\alpha $ \cite{Lee2013a}. 
Although the preparation of these resource states is non-deterministic (success probabilities 1/4 and 1/16, respectively), it is the only non-deterministic  off-line process after hybrid pairs are prepared.
Due to the near-deterministic nature of HBSM, the scheme already showed advantages over previous ones such as
CSQC \cite{Lund2008a} and parity-state-based LOQC \cite{Hayes2010a}
in terms of both loss thresholds and resource requirements at an encoding amplitude $ \alpha_{\rm opt} \approx 1.08 $ under the 7-qubit Steane code.

\begin{figure}[t]
\includegraphics[width=.9\textwidth]{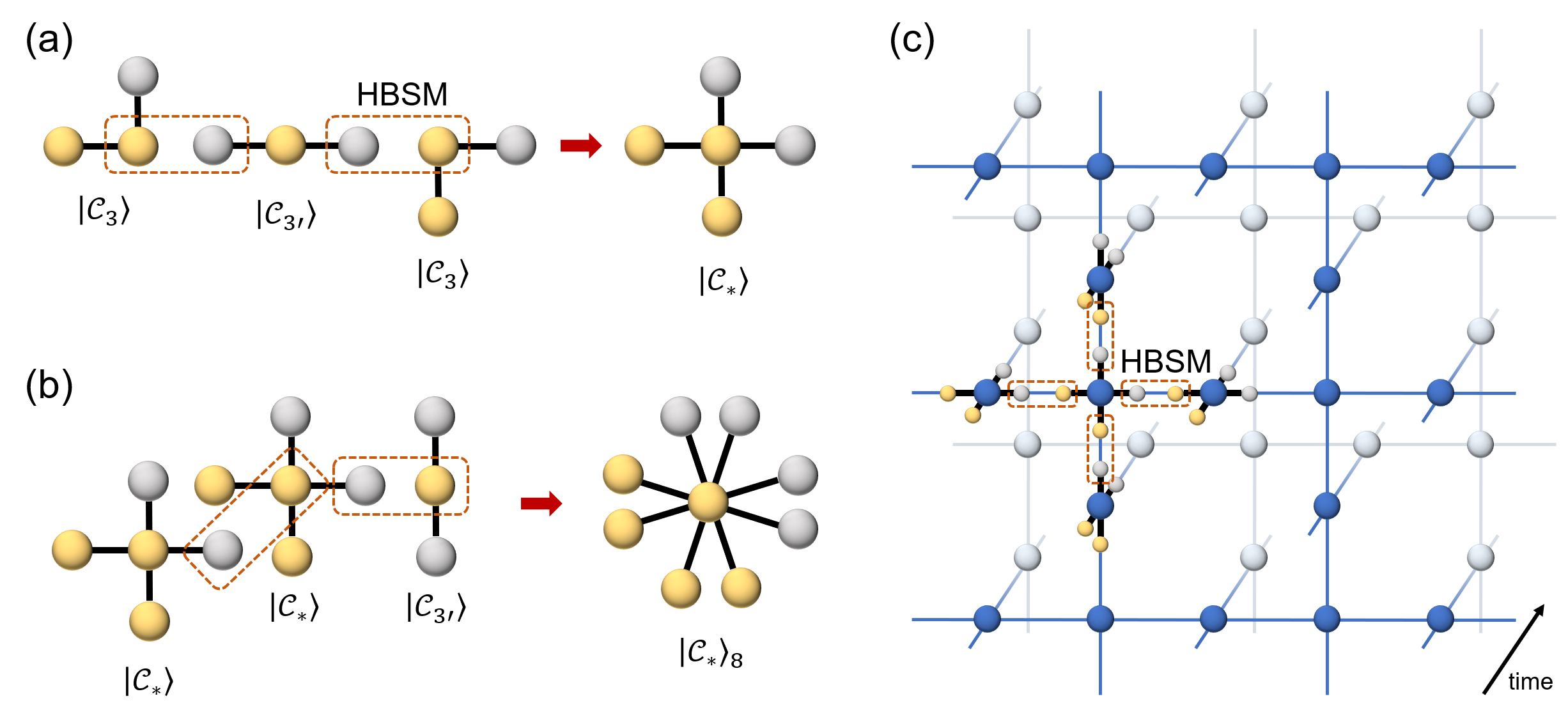}
    \centering
\caption{
Major steps for hybrid quantum computing. 
(a) In HTQC, three-qubit cluster states, two $\ket{\mathcal{C}_3}$'s and one $\ket{\mathcal{C}_{3^\prime}}$, are fused using HBSMs to ballistically form $\ket{\mathcal{C}_\ast}$. (b) In PHTQC, two $\ket{\mathcal{C}_\ast}$'s are fused along with a $\ket{\mathcal{C}_{3^\prime}}$ to form a $\ket{\mathcal{C}_\ast}_8$ and the process can be extended to form a $\ket{\mathcal{C}_\ast}_{4n}$, where $n>2$. Here, the states are post-selected over all BSMs being successful. (c) Physical implementation of HTQC and PHTQC-$n$ requires two layers of RHG lattice to be present and the other layers are formed as the FTQC progresses by layer-by-layer measurement of qubits in $X$-basis, $M_X$. The layers are formed and interconnected by placing $\ket{\mathcal{C}_\ast}$ ($\ket{\mathcal{C}_\ast}_{4n}$) on the nodes forming edges by performing HBSMs ($n$ repeated HBSMs).  }
\label{fig:HTQC}
\end{figure}

The performance of hybrid quantum computation can be enhanced by adopting topological QEC codes.
Hybrid-qubit-based topological quantum computation (HTQC)~\cite{Omkar2020a} uses hybrid qubits to generate the 
Raussendorf-Harrington-Goyal (RHG) lattice~\cite{Raussendorf2006a}. The protocol begins by generating two types of three-qubit cluster states, $\ket{\mathcal{C}_3}  \propto \ket{0_\textrm{L},0_\textrm{L},0_\textrm{L}} + \ket{0_\textrm{L},0_\textrm{L},1_\textrm{L}} + \ket{1_\textrm{L},1_\textrm{L},0_\textrm{L}} - \ket{1_\textrm{L},1_\textrm{L},1_\textrm{L}} $ and $\ket{\mathcal{C}_{3^\prime}}\propto \ket{0_\textrm{L},0_\textrm{L},0_\textrm{L}} + \ket{1_\textrm{L},1_\textrm{L},1_\textrm{L}}$, using $\textrm{B}_{\rm I}$ and $\textrm{B}_\alpha$ on hybrid pairs with a success probability of $\sim 1/2$, respectively~\cite{Omkar2020a}.
The RHG lattice is then ballistically constructed in two steps: first, three-qubit cluster states are fused to create 5-qubit star cluster states $ \ket{\mathcal{C}_\ast} $ as shown in Fig.~\ref{fig:HTQC}(a); these are subsequently arranged on the lattice and further fused to complete the RHG structure [see Fig.~\ref{fig:HTQC}(b)]. The loss threshold of HTQC is 
$ \eta_\textrm{th} \approx 5.07 \times 10^{-3} $, 
which is an order higher than that of HQQC, while the encoding amplitude 
$ \alpha_{\rm opt} \approx 1.5 $ is slightly larger. HTQC achieves a substantial reduction in resource requirements by approximately four orders of magnitude compared to HQQC.

A post-selected HTQC (PHTQC-$n$) \cite{Omkar2021a} employs a smaller encoding amplitude $\alpha$ to mitigate dephasing errors caused by photon loss, while compensating for the resulting reduction in the BSM success probability by $ n $ repeated BSMs. This approach utilizes star cluster states with $ 4n $ arms, $ \ket{\mathcal{C}_\ast}_{4n} $, and they are fused by multiple attempts of BSMs to create entanglement along edges of RHG lattice. The state $ \ket{\mathcal{C}_\ast}_{4n} $ can be generated by merging three-qubit cluster states as shown in Fig.~\ref{fig:HTQC}(c). Table~\ref{tab:para} shows that PHTQC-2 and PHTQC-3 achieve highest loss thresholds over 1\%. Moreover, the optimal encoding amplitude is $\alpha_{\rm opt}\approx 0.84$ ($\alpha_{\rm opt}\approx0.6$) for PHTQC-2 (PHTQC-3), clearly demonstrating the relaxed requirement for state preparation. Note that for $ n>3 $, PHTQC-$n$ does not take advantage in reducing the resource overhead, because $\alpha_{\rm opt}$ becomes even smaller, making the failure probability of HBSM comparable to that of the DV scheme \cite{Lee2013a, Lee2015a}.

By employing the bosonic error correction code in the bosonic encoding, the noise resilience of the hybrid qubit may be further enhanced. Recently, a hybrid-cat-code quantum computation (HCQC) scheme was proposed \cite{Lee2024a}, employing the four-headed cat code,
where a qubit is defined as $ a\ket{+}(\ket{\alpha}+\ket{-\alpha}) + b\ket{-}(\ket{i\alpha}+\ket{-i\alpha}) $. The BSM in the basis of four-headed cat code \cite{Hastrup2022a, Su2022a} can distinguish all four Bell states near-deterministically even in the case of single photon loss, so that the errors by single photon loss can be corrected. The basic building block of this scheme is a hybrid-cat pair $ \ket{H}(\ket{\alpha}+\ket{-\alpha}) + \ket{V}(\ket{i\alpha}+\ket{-i\alpha}) $, which can be generated by extending the components of a cat state using the scheme of Ref. \cite{Hastrup2020a} with probability of $\approx1/8$ \cite{Lee2024a}. For the sake of error-correcting capabilities of four-headed cat code, 
the number of required hybrid states are significantly reduced as seen in Table~\ref{tab:para} at the cost of requiring a larger encoding amplitude $ \alpha_{\rm opt}\approx 2.93 $.

We note that 
the threshold analysis in Table~\ref{tab:para} varies by error model employed in each scheme.
In HTQC  and PHTQC-$n$, the BSM failures correspond to missing edges that are more tolerable than $Z$ errors in the RHG lattice~\cite{Auger2018a}. Furthermore, HTQC employs adaptive single-qubit-measurements to improve the error-correcting ability of the lattice~\cite{Auger2018a}. 
Rather, in HCQC \cite{Lee2024a}, the ambiguity in HBSM is treated as a logical $X$ error and its propagation is thoroughly analyzed.
Also, in HQQC \cite{Lee2013a} based on telecorrection, only losses in the teleported state were considered during the HBSM, assuming lossless teleportation resource states, whereas later works \cite{Omkar2020a,Omkar2021a,Lee2024a} consider losses in both modes of the HBSM.

\subsection*{Generation of hybrid entanglement}
\begin{figure}[t]
    \centering
    \includegraphics[width=0.95\columnwidth]{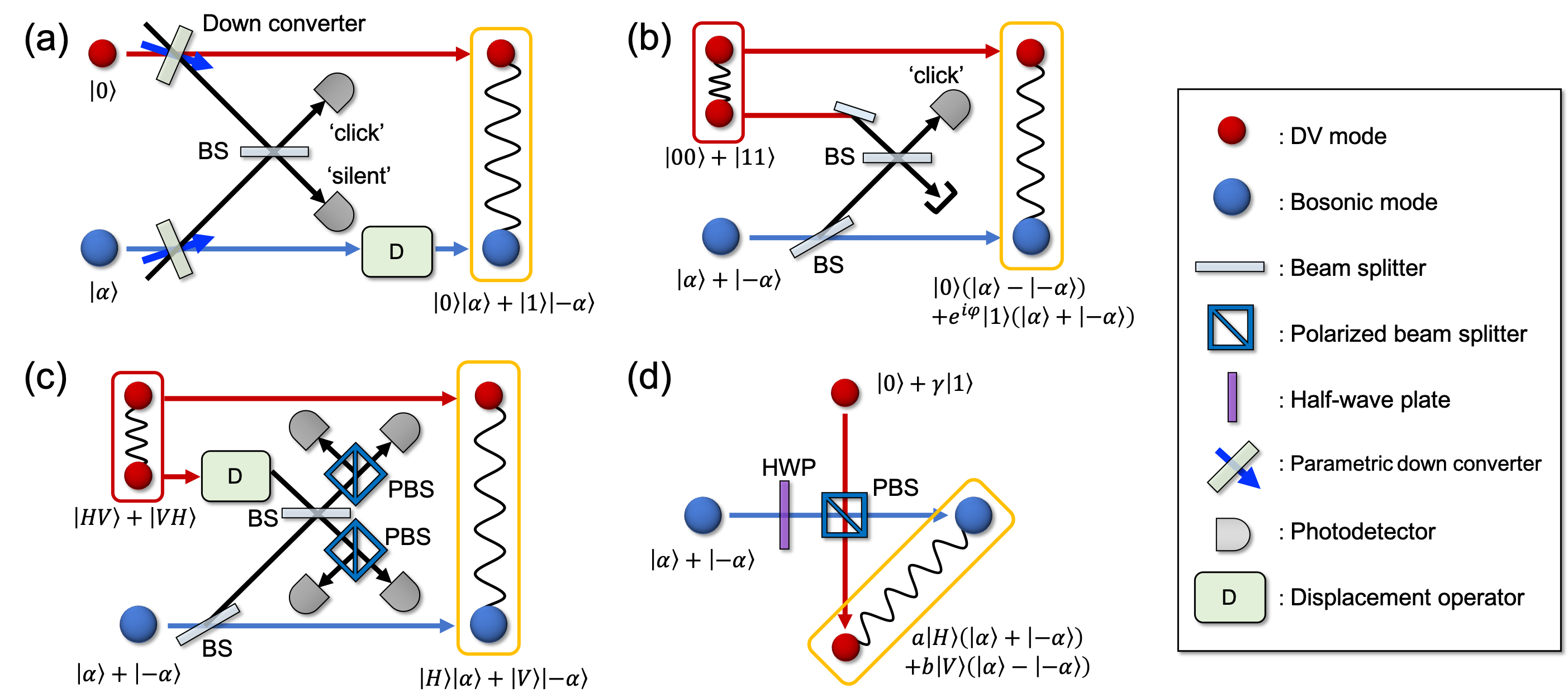}
    \caption{Schemes for generating photonic hybrid entanglement. Single-rail hybrid entanglement is generated via (a) conditional photon addition~\cite{Jeong2014a} and (b) photon subtraction~\cite{Morin2014a}. Dual-rail hybrid entanglement is generated via (c) heralded photon detection after a DV-entangled state and a cat state pass through a beam splitter~\cite{Kwon2015}, and (d) coherent mixing of a cat state and superposed DV modes through a polarizing beam splitter~\cite{Sychev2018a}.}
    \label{fig:Hybrid_Generation}
\end{figure}

The hybrid FTQC schemes employ hybrid entangled pairs $ \ket{H}\ket{\alpha} + \ket{V}\ket{-\alpha} $ as basic building blocks and such resources are available in laboratories \cite{Jeong2014a, Morin2014a, Ulanov2017a,  Sychev2018a, Guccione2020a, Darras2023a} [see Fig.~\ref{fig:Hybrid_Generation}]. A hybrid pair combining a single-rail (vacuum and single-photon) qubit and a bosonic qubit can be generated via a conditional photon addition or subtraction operation on the bosonic mode, in a manner similar to the generation of a cat state. In the scheme of Ref. \cite{Jeong2014a}, a photon is added via parametric downconversion either to the DV mode or to the bosonic mode, where the idler photon is detected heralding the erasure of which-path information [see Fig.~\ref{fig:Hybrid_Generation}(a)].
On the other hand, in the scheme of Ref. \cite{Morin2014a}, a photon is subtracted probabilistically from the bosonic mode, and it interferes with the DV mode to herald the generation of a hybrid entangled state [see Fig.~\ref{fig:Hybrid_Generation}(b)]. For the generation of dual-rail hybrid entanglement, a heralded detection protocol was theoretically proposed in Ref.~\cite{Kwon2015} [see Fig.~\ref{fig:Hybrid_Generation}(c)], and a different approach employing a dual-rail qubit and a bosonic qubit was experimentally realized more recently~\cite{Sychev2018a} [see Fig.~\ref{fig:Hybrid_Generation}(d)]. With such experimental developments, an encoding amplitude $ \alpha \approx 0.9 $ has been achieved \cite{Sychev2018a, Darras2023a}, which is almost within reach of realizing practical hybrid quantum computation.
Various implementations of hybrid entangled states~\cite{Jeong2014a, Morin2014a, Ulanov2017a, Sychev2018a, Guccione2020a, Darras2023a} offer a feasible route for employing hybrid qubits as logical qubits. This provides an advantage of the hybrid approach over other bosonic quantum computing schemes such as GKP- and binomial-state-based ones that require high levels of squeezing or strong nonlinear interactions.

\subsection*{Hybrid quantum networks}

\begin{figure}[t]
\includegraphics[width=0.8\columnwidth]{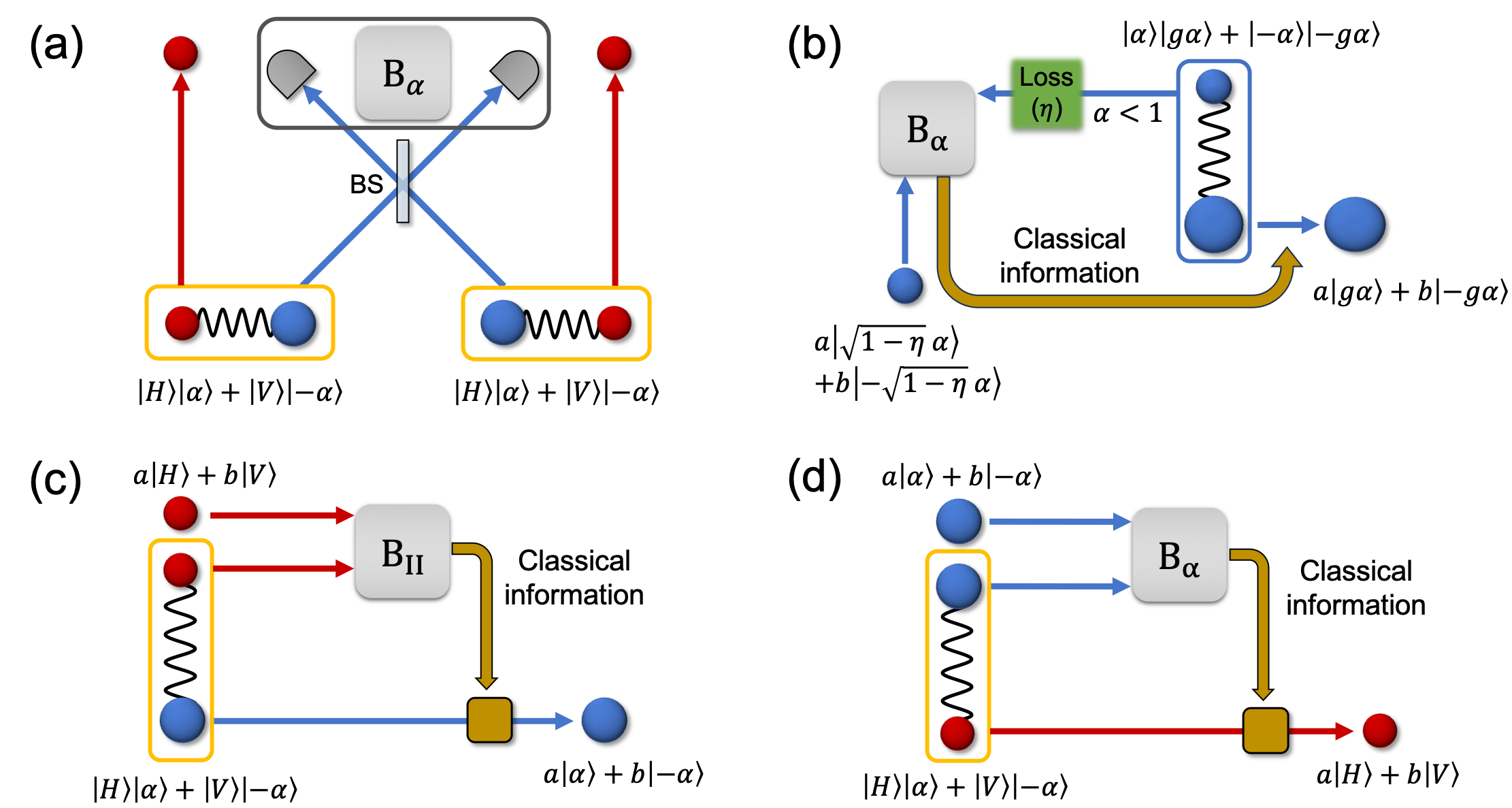}
        \centering
	\caption{\label{fig:QN}
    (a) Loss-resilient entanglement swapping using  hybrid pairs and $\textrm{B}_\alpha$ \cite{Lim2016a}. Here, $\textrm{B}_\alpha$ can be displaced with other measurements strategies such as homodyne detection for better performance depending on the loss environment \cite{Lim2016a}. 
    (b) Tele-amplification of a cat-state qubit for loss-resilient communication \cite{NeergaardNielsen2013}.
    Teleporation  (c) from a single-photon qubit to a cat-state qubit \cite{Sychev2018a,Darras2023a} and (d) from a cat-sate to a single-photon qubit \cite{Ulanov2017a}. }
\end{figure}

When quantum computing is integrated with quantum networks, it can unlock the true potential of future technologies such as the quantum internet, distributed quantum algorithms, and quantum cloud computing. 
There have been studies on entanglement swapping~\cite{Lim2016a,Guccione2020a}, quantum repeaters~\cite{vanLoock2006a, Brask2010a, Sheng2013a, Lim2016a, Guccione2020a}, and quantum teleportation protocols~\cite{Park2012a,Jeong2016a,Kim2016a,Choi2020a,Bose2022a,Bera2025a,Takeda2013a,Ulanov2017a,Sychev2018a,Darras2023a} in the context of photonic hybrid architectures, which will serve as bridging units that support quantum networks. 
Advancements in quantum-repeater primitives that directly manipulate hybrid entangled states have paved the way to the design of hybrid repeaters that can truly distribute these states. First, linear-optical-based entanglement purification of (multiphoton and/or multicoherent) hybrid entangled states that can purify bit-flip, phase-flip and coherent-state photon-loss errors was introduced~\cite{Sheng2013a}. Loss-tolerant hybrid entanglement swapping was also proposed~\cite{Lim2016a} [see Fig. \ref{fig:QN}(a)]. Recently, the distribution of hybrid entangled resource states has been theoretically shown to be achievable at distances as far as 300~km through photon-lossy telecommunication optical fibers just with one round of hybrid entanglement swapping midway the transmission~\cite{Bose2024a}.

There are at least two reasons why entanglement swapping and quantum repeaters using hybrid states may offer significant advantages. First, as previously discussed, hybrid-state methods can perform highly efficient BSMs. 
Entanglement swapping relies on BSMs, and the hybrid-state-based approach allows for efficient BSM by utilizing the coherent-state part. 
Second, by adjusting the amplitude of the coherent-state component in hybrid states, one can construct a loss-tolerant architecture suitable for long-distance transmission as shown in Fig.~\ref{fig:QN}(b). This advantage has already been experimentally demonstrated through tele-amplification \cite{NeergaardNielsen2013}, wherein a state with a small coherent amplitude is transmitted over a long distance but the amplitude is amplified after transmission.
Furthermore, the qubit-converting technology between coherent-state qubits and single-photon qubits has not only been theoretically explored using the teleportation protocol \cite{Jeong2016a,Choi2020a} but also experimentally demonstrated \cite{Ulanov2017a,Sychev2018a,Darras2023a} [see Fig. \ref{fig:QN}(c,d)]. It allows one to covert a cat-state qubit to a single-photon qubit and vice versa. These techniques will be powerful tools for constructing hybrid networks, especially since the operations that can be efficiently performed by the coherent-state and single-photon parts may differ.

\section*{Remarks and future prospects}

The photonic hybrid quantum computing utilizing cat-state qubits and single-photon qubits \cite{Lee2013a, Omkar2020a, Omkar2021a, Lee2024a} combines the advantages of both modalities while significantly overcoming the limitations of conventional methods, and offers substantial benefits for the realization of fault-tolerant quantum computing. Specifically, this hybrid scheme enables nearly-deterministic universal gate operations, thereby greatly enhancing resource efficiency. At the same time, it allows for ballistic MBQC without active feedforward, thereby overcoming the limitations of single-photon-qubit-based approaches.
Recent proposals based on single-photon qubits such as MBQC \cite{Omkar2022a,Lee2023a} and FBQC \cite{Bartolucci2023a,Song2024a}, while capable of achieving high error thresholds, require either active feedforwards \cite{Omkar2022a,Lee2023a,Song2024a} or excessive resources \cite{Bartolucci2023a,Song2024a,Bartolucci2025}. Thus, achieving both high resource efficiency and ``active-feedforward-free" is a significant practical advantage of the hybrid approach. Maintaining this combination of high resource efficiency and the ballistic nature, while still offering a reasonable loss threshold around 1\% per mode (i.e., $\approx 2\%$ per hybrid qubit), makes the hybrid approach  a promising candidate for implementing photonic quantum computing.

Another strong contender in photonic quantum computing is the GKP-state-based approach
\cite{Gottesman2001a}, which offers advantages in error correction efficiency. However, it faces a major challenge of generating the photonic GKP states themselves. In contrast, the hybrid approach benefits from the fact that its fundamental resource -- namely, the hybrid qubit -- has known implementation methods and has already been demonstrated experimentally \cite{Jeong2014a, Morin2014a, Sychev2018a}.

We have compared several hybrid schemes proposed to date. Taking into account the values of $\eta_{\rm th}$, $N$, and $\alpha_{\rm opt}$ in Table~\ref{tab:para}, PHTQC-2 \cite{Omkar2021a} might be the most advantageous among the suggested schemes. HCQC \cite{Lee2024a} could work in a more resource-efficient way if improved techniques to generate large-amplitude ($\alpha\approx2.9$) hybrid pairs are developed. Further advancements may be possible through integration with techniques such as low-density parity-check codes \cite{MacKay2004}, FBQC \cite{Bartolucci2023a}, and autonomous error correction \cite{Leghtas2013a}, as well as through the use of single-rail logic \cite{Lund2002a} for the DV mode as its merit for teleportation was shown \cite{Kim2016a}.

Beyond photonics, hybrid architectures in circuit‑QED and trapped‑ion platforms harness the enhanced non‑linearity of bosonic modes to efficiently implement bosonic gates~\cite{Kudra2022a}, prepare  non-classical states~\cite{Kudra2022a, Eickbusch2022a}, and scale up the number of qubits~\cite{Kang2025}.
Indeed, cross-system interactions in hybrid systems enable autonomous error correction, demonstrated by long‑lived GKP and cat codes stabilized via dispersive or two‑photon‑dissipation processes \cite{Sivak2023a, Lachance-Quirion2024a}. It is an open challenge to engineer such mechanisms for photonic hybrid qubits. 

Scalable and fault-tolerant quantum computing requires extreme control of quantum states. Different physical degrees of freedom have different advantages and challenges. In order to achieve required control of quantum states,
a hybrid approach is a natural direction to pursue. In this regard, manipulating the interfaces in hybrid systems will be an important issue.

\section*{ACKNOWLEDGMENTS}

We are grateful to Seung-Woo Lee for valuable discussions and suggestions. 
J.L. is supported by the Korea Institute of Science and Technology (2E33546, 2E33541) and the National Research Foundation (NRF) of Korea (2022M3K4A1094774, RS-2024-00509800).
S.O. is supported by the National Research Foundation of Korea (NRF) grant funded by the Korea government (MSIT) (Nos. RS-2024-00413957).
Y.S.T. is  supported by National Research Foundation (NRF) of Korea (RS-2023-00237959, RS-2024-00413957, RS-2024-00437191, RS-2024-00438415 and RS-2025-02219034).
S.H.L. is supported by the Australian Research Council via the Centre of Excellence in Engineered Quantum Systems (EQUS) Project No. CE170100009.
H.K. is supported by the KIAS Individual Grant No. CG085302 at Korea Institute for Advanced Study.
M.S.K. acknowledges funding from the UK EPSRC through EP/Z53318X/1, EP/W032643/1 and EP/Y004752/1, the KIST through the Open Innovation fund and the National Research Foundation of Korea grant funded by the Korean government (MSIT) (No. RS-2024-00413957).
H.J. is supported by the National Research Foundation of Korea (NRF) grant funded by the Korea government (MSIT) (Nos. RS-2024-00413957, RS-2024-00438415, and NRF-2023R1A2C1006115) and by the Institute of Information \& Communications Technology Planning \& Evaluation (IITP) grants funded by the Korea government (MSIT) (IITP-2025-RS-2020-II201606 and IITP-2025-RS-2024-00437191).

\bibliography{ReviewHybridQC}

\end{document}